\documentclass[a4paper,twocolumn]{PoS}

\usepackage[utf8]{inputenc}
\usepackage{multicol}
\usepackage{enumerate}
\usepackage{upgreek}
\usepackage{lineno}
\usepackage{amssymb}
\usepackage{array}
\usepackage{amsmath}

\title{Challenges for the Adaptive Gain Integrating Pixel Detector (AGIPD) design due to the high intensity photon radiation environment at the European XFEL}

\ShortTitle{AGIPD design challenges due to the high intensity photon radiation environment at the European XFEL}

\author{\speaker{J. Becker}, L. Bianco, P. Göttlicher, H. Graafsma, H. Hirsemann, S. Jack, A. Klyuev, S. Lange A. Marras, U. Trunk\\
        DESY, Hamburg, Germany\\
        E-mail: \email{Julian.Becker@desy.de}}

\author{R. Klanner, J. Schwandt, J. Zhang\\
	University of Hamburg, Germany\\
        }

\author{R. Dinapoli, D. Greiffenberg, B. Henrich, A. Mozzanica, B. Schmitt, X. Shi\\
	PSI, Villigen, Switzerland\\
       }

\author{M. Gronewald, H. Krüger\\
	University of Bonn, Germany\\
       }

\abstract{

The European X-ray Free Electron Laser (XFEL) is a new research facility currently under construction in
Hamburg, Germany. With a pulse length of less than 100 fs and an extremely high
luminosity of 27000 flashes per second the European XFEL will have a unique time structure
that demands the development of new
detectors tailored to the requirements imposed by the experiments while complying with the machine
specific operation parameters. The Adaptive Gain Integrating Pixel Detector (AGIPD) is one response to
the need for large 2D detectors, able to cope with the 4.5 MHz frame rate, as well as with the high
dynamic range needed by XFEL experiments ranging from single photons to more than 10$^4$ 12 keV photons per pixel per
pulse. In addition it has to withstand doses of up to 1 GGy over three years.

}

\FullConference{The 21st International Workshop on Vertex Detectors\\
		16-21 September 2012\\
		Jeju, Korea}

\begin{document}

\setlength{\oddsidemargin}{1.5cm} 
\setlength{\evensidemargin}{1.5cm}
\setlength{\textwidth}{18cm}
\setlength{\columnsep}{20pt}
\twocolumn

\section{Introduction}

The European X-ray Free Electron Laser (XFEL) \cite{XFEL, TN2011-001} will provide ultra short, highly coherent X-ray pulses which will revolutionize scientific experiments in a variety of disciplines spanning physics, chemistry, materials science, and biology. 

Dedicated fast 2D detectors for the European XFEL are being developed, one of which is the Adaptive Gain Integrating Pixel Detector (AGIPD) \cite{AGIPD1, AGIPD2, AGIPD3}, developed by a collaboration between DESY, the University of Hamburg, the University of Bonn (all in Germany) and the Paul Scherrer Institute (PSI) in Switzerland.

\section{Requirements}

One of the differences between the European XFEL and other free electron laser sources is the high pulse repetition frequency of 4.5~MHz. The European XFEL will provide pulse trains, consisting of up to 2700 x-ray pulses of less than 100~fs duration, separated by 220~ns (600~$\upmu$s in total), followed by an idle time of 99.4~ms, resulting in a supercycle of 10~Hz and 27000 pulses per second. The energy of the x-rays will be tunable in a range depending on the experimental station. Together the beamlines will cover the energy range from a few hundreds of eV to several tens of keV.

As XFELs develop an enormous power density at the sample location, only very few samples will survive illumination. Most samples will be destroyed by the x-ray pulse \cite{explosion}, making the acquisition of the diffracted patterns from a single pulse mandatory. Obviously it is desirable to acquire as many of the 2700 images per train as possible.

As the dynamic range required by common experiments is very large, from many thousands of photons close to the central beam to essentially single photon events at large angles, the dynamic range of the employed detection system has to be large as well.

It should be noted that not only the high number of photons close to the central beam and in Bragg spots needs to be measured correctly, but the reliable detection of single photons and their discrimination against background is equally, if not even more, important.

Additionally, all detector systems for the European XFEL need to be sufficiently radiation hard to survive an estimated dose of 10$^{16}$ 12.4~keV photons, corresponding to 1 GGy at the entry window \cite{dose}, during 3 years of operation.

\section{The Adaptive Gain Integrating Pixel Detector (AGIPD)}

AGIPD is based on hybrid pixel technology and aims at imaging in the energy range between 3 and 15~keV. The current design goals of the newly developed Application Specific Integrated Circuit (ASIC) with independent dynamic gain switching amplifiers in each pixel are (for each pixel) a dynamic range of more than 10$^4$ 12.4~keV photons in the lowest gain, single photon sensitivity in the highest gain, and operation at 4.5~MHz frame rate. An external veto signal can be provided to maximize the number of useful images by overwriting any image previously recorded during the pulse train. 

Due to the special pulse structure of the European XFEL, it is necessary to store the acquired images inside the pixel circuit area during the pulse train. A compromise had to be found between storing many images, requiring a large pixel area, and high spatial resolution, requiring small pixel sizes \cite{AGIPD1, AGIPD2}. The image data is read out and digitized in the 99.4~ms between pulse trains.

The AGIPD will feature a pixel size of (200~$\upmu$m)$^2$, which is sufficient to accommodate an analog memory for 352 images. For many experiments using particle injection mechanisms the memory of AGIPD should be sufficient, as particle hit rates are currently below 10\% (see \cite{cdr_spb} and references therein). The impact of the limited number of storage cells on X-ray Photon Correlation Spectroscopy (XPCS) as intended to be used on the MID station \cite{cdr_mid} depends on the properties of the sample and has been investigated in \cite{xpcs_sampling}.

\subsection{Sensor}

\begin{table*}
	\centering	\begin{tabular}{c|c|c}
			\textbf{Property}		& \textbf{Specification} 	& \textbf{Comment} \\
			\hline
			\textbf{readout electrodes}		& p$^+$ implants 			&  hole collection\\ 
			\textbf{bulk doping}		& 3-8 k$\Omega\cdot$cm 			&  n-type bulk material\\ 
			\textbf{dimension}		& 107.6 mm $\times$ 28 mm &  large monolithic sensors reduce the overall dead area\\ 
			\textbf{thickness}	& 500~$\upmu$m 			& > 90\% quantum efficiency at 12.4~keV \\
			\textbf{effective entry window thickness}& $<$ 2.5~ $\upmu$m & important for QE at lower energies \\
			\textbf{bias voltage} & $>$ 500 V & minimizes impact of plasma effects
														
		\end{tabular}
		\caption{Important major specifications of the AGIPD sensor.}
		\label{sensor_specs}
\end{table*}

The design of the AGIPD sensor is especially challenging, as it needs to be very radiation hard (approximately 90\% of the total dose is absorbed in the sensor) and it needs to be biased with a high voltage of 500~V or more, the reasons for which are explained below. 

The sensor will be made of silicon, exploiting the wide experience with radiation damage in silicon and the high number of vendors able to manufacture specialized designs. The most important design parameters are shown in table \ref{sensor_specs}. In order to provide a sufficiently high quantum efficiency (QE) for photons of 12.4~keV energy and above, a sensor with a thickness of 500~$\upmu$m was chosen. At the same time the entry window was carefully designed to maintain a significant QE for energies of 3~keV and below. Details on the sensor and pixel layout can be found in \cite{schwandt, schwandt2}.

The effect of extreme doses of ionizing radiation on typical sensor structures was investigated in great detail \cite{rad_dam,jiaguo1,jiaguo2}, and, using simulations which take into account the effects of radiation damage, a sensor layout, featuring narrow gaps between pixel implants and an optimized guard ring, was developed that should be able to stand a sufficiently high voltage after irradiation \cite{schwandt}.

The reason for biasing with such a high voltage is the plasma effect, which is produced by localized high instantaneous charge densities, which in turn are created by thousands of photons arriving simultaneously. The impact of the plasma effect manifests mainly in increased charge spreading\footnote{The FWHM of the charge cloud increases by a factor of $> 3$ at intensities $> 10^4$ 12.4~keV photons and low applied voltages.} and increased charge collection times\footnote{The time to collect 95\% of the charge increases by a factor of $> 4$ at intensities $> 10^4$ 12.4~keV photons and low applied voltages.}, the details of which have been studied in \cite{plasma, thesis, plasma2}. It was found that a high sensor bias voltage suppresses the impact on the scientific imaging. It is expected that plasma effects will manifest themselves in the parts of diffraction patterns close to the central beam and the HORUS simulation tool \cite{horus1,horus2,horus3} was developed to evaluate the scientific impact.

\subsection{Application Specific Integrated Circuit (ASIC)}

The AGIPD readout ASICs will be manufactured in IBM 130 nm CMOS technology. Several small (16x16 pixels) prototypes have been produced using Multi Project Wafer (MPW) runs, both to test the technology characteristics and to evaluate the best architectural solutions to be employed \cite{trunk}. Radiation-hard design techniques are employed, including the use of Enclosed Layout Transistors (ELTs) and guard rings around critical devices.

Design issues for the full scale (64x64 pixels) chip are settled, and a submission to the foundry is scheduled beginning of 2013.

\begin{figure*}[tb!]
  \includegraphics[width=0.95\textwidth]{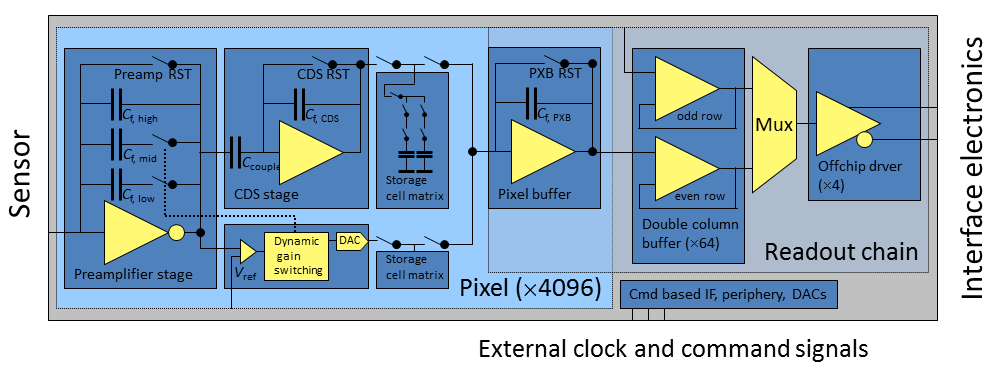}
  \centering
  \caption{Block diagram of the circuitry on the AGIPD 1.0 chip. Reset switches are abbreviated RST and the Correlated Double Sampling stage CDS. The double column buffer works in an interleaved way by connecting only to odd or even rows in each column, thereby relaxing the timing constraints on the buffer.}
  \label{block}
\end{figure*}

A block diagram of the full scale (64x64 pixels) chip AGIPD 1.0 is shown in figure \ref{block}. The typical signal path comprises charge generation and transport within the sensor, charge collection and (amplified) charge to voltage conversion in the adaptive gain amplifier, Correlated Double Sampling (CDS) of the amplifier output voltage by the CDS stage, signal storage in the analog memory cells, readout of the storage cells and transfer of the signal to the outside world via differential lines for the analog signals. Commands and clock signals are received via LVDS lines.

\subsubsection{Noise performance}
For pristine chips under standard operation conditions\footnote{Standard operating conditions employ about 100~ns integration time. For such short exposure times the contribution of leakage currents, even when elevated after irradiation, is negligible.} the amplifier, CDS stage and readout contribute about equally to the total noise, which for any given sample follows a Gaussian distribution\footnote{Although all observations support it, this assumption is not trivial, especially for tails more than 5 sigma away from the mean (i.e. very rare events).}. Therefore it is convenient to express the noise in terms of the rms of the Equivalent Noise Charge (ENC). A more detailed noise analysis has been presented before \cite{AGIPD2, iworid, noise_paper}.



Measurements from the most recent test chip indicate that for AGIPD 1.0 a noise of approximately 300 electrons will be reached \cite{noise_dominic}. The precise value of the noise of the full scale chip is only known approximatively at the moment. Assuming that the results from the test chip are mostly transferable to AGIPD 1.0, the  ENC of the full scale chip, featuring 64~$\times$~64 pixels, can be estimated to be between 200 and 400 electrons.


\subsubsection{Radiation damage}

%
Certain parts of the electronics will be exposed to significant amounts of ionizing radiation. Even when taking the shielding of the sensor into account, it is expected that some parts of certain ASICs may accumulate up to 100~MGy of dose during their expected operation time of 3 years.

In order to investigate the effect of these high levels of radiation an irradiation campaign has been performed using the DORIS F4 irradiation facility at DESY and prototypes (AGIPD 0.3, AGIPD 0.4) were irradiated to total doses from 1 to 100~MGy.

The test chips have been found to tolerate doses up to at least 10~MGy. After irradiation the performance of irradiated chips up to this dose is comparable to that of pristine ones, although an increase in noise between 30\% and 40\% was noted \cite{pixel2012}, which seems to saturate above approximately 1~MGy.  Further irradiation to 100~MGy rendered the chip non-functional\footnote{Tentatively this happens due to charge accumulation in the oxide layer, which causes a substantial shift of the threshold voltages of the devices.}. After thermal annealing the functionality of the chip is recovered, albeit with a reduced analog performance which is still under investigation. It should be noted that the irradiation of the chip happened in an accelerated fashion compared to operating conditions (accumulating 100~MGy of dose required about 1 week of irradiation at room temperature), so it can be speculated that the thermal annealing during maintenance periods and when the final detector system is not in use might be sufficient to prevent the chip from losing its functionality.

\begin{figure}[tb!]
  \includegraphics[width=0.45\textwidth]{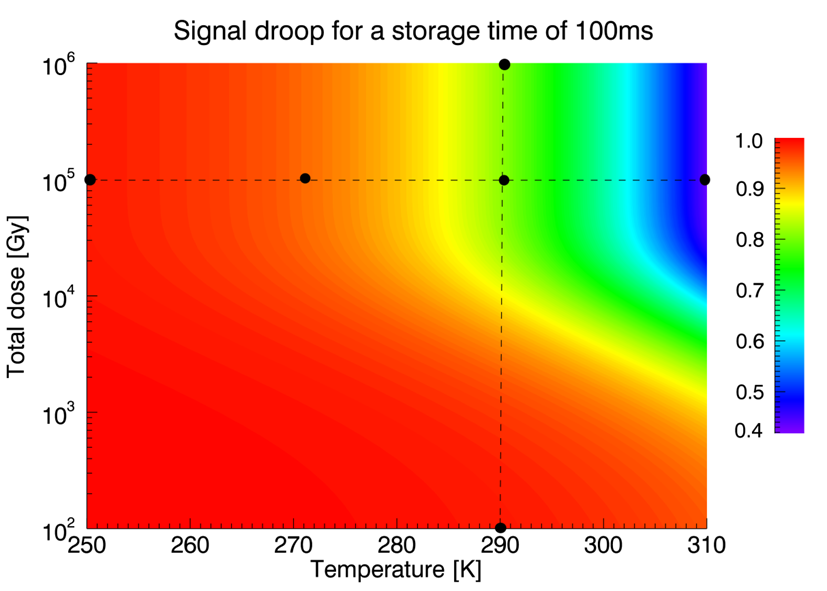}
  \centering
  \caption{2D representation of the effect of droop (fraction of charge remaining in the storage cell after 100~ms wait time) as function of dose and temperature. The measurements (dark points) were presented in \cite{droop}. Typical storage times for the analog information will be below 12~ms. The data point at 100~Gy corresponds to the situation before irradiation.}
  \label{2d_droop}
\end{figure}

Additionally it was found that after heavy irradiation and/or at elevated temperatures the signal droop of the storage cells becomes non negligible. Reducing the operating temperature has been shown to mitigate this effect \cite{droop}, and detailed investigations are currently ongoing. In response to these results it was decided to reduce the operating temperature to -20$^\circ$~C, at which the droop becomes almost negligible even after irradiation (>99\% of signal remaining) . A 2D overview showing the measured droop at different temperatures and doses for a storage time of 100~ms is shown in Figure \ref{2d_droop}. The regions not covered by measurement points were extrapolated assuming the influence of temperature and dose are independent of each other, described by exponential functions with a constant offset and their effect is multiplicative. An elaborate measurement campaign to investigate the behavior of the droop in detail is currently ongoing.

\subsection{Interface electronics}

The interface electronics comprises of all the electronics between the ASICs, which are specially developed for AGIPD, and the data acquisition system, which is a common development for all detector systems at the European XFEL.

It can be roughly subdivided into two parts: the control system and the read-out system, both of which are detailed below.

In the mechanical layout, which is detailed later on, care was taken to place as much of the interface electronics as possible outside of the detector vacuum. In this way a forced air stream cooling concept can be realized, which minimizes bulky mechanics outside the detector vacuum and increases the serviceability of electronic components.

\subsubsection{Control system to operate the ASIC and select the best bunches (veto)}

The external control system hardware has been developed using MTCA.4 crate standards which are common to all European XFEL detectors \cite{cook}. An FPGA in the control part of the camera head receives control line signals, performs bookkeeping of free storage cells, and instructs the ASICs which storage cell to use next. The bookkeeping information will be added to the main data stream. In this way it is known to which storage cell a bunch was assigned for later off-line data processing.

An intelligent power supply outside the experimental hutch delivers the power for the ASICs, approximately 500~A at 1.5~V, and interface electronics and the high voltage of 500~V or more to the sensors. The control system communication will be based on 10/100~Mb Ethernet using the TCP/IP protocol and will connect both the camera head and the power system. Within the detector head the (slow) control information is distributed and collected by a multi-branch I$^2$C network.

There is an additional PCB board in the vacuum connecting the electronics outside of the vacuum to the modules inside. This vacuum board's purpose is the routing of the analog signal lines to the boards described above and the regulation of the voltage on the power lines for the ASICs.

\subsubsection{Analog and digital read-out electronics}

\begin{figure*}[tb!]
  \includegraphics[width=0.9\textwidth]{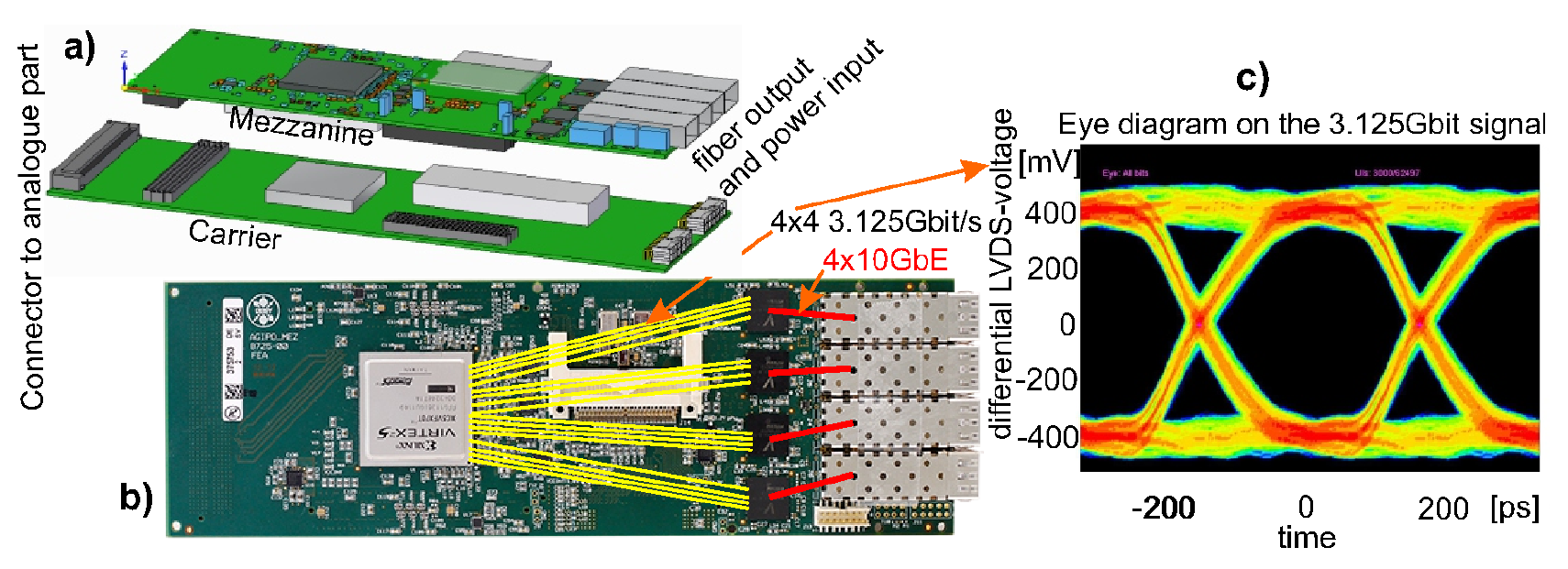}
  \centering
  \caption{Digital PCB assembly for a module: (a) the mechanical concept, (b) the mezzanine card (c) the eye diagram of the high speed interconnect (3.125 Gbit/s). Image reproduced from \cite{zimmer}.}
  \label{digital_pcb}
\end{figure*}


Directly outside the vacuum vessel, a system of analog PCBs perform line reception, pickup noise filtering and digitization of the analog signals from the ASICs. A single stage filter suppresses the noise at the 33~MHz sampling frequency by 3~dB and settles to better than 11~bits accuracy. The analog system also drives the digitized signal to a stack of digital boards which then handle the communication with the outside world. 

The signal of two ASICs is digitized by one AD9257\footnote{Each AD9257 component contains 8 ADC circuits.} fast serial output ADC, which is commonly used in medical imaging applications. The data rate per ADC is 64x64/4x352x14x2x10 bits/s\footnote{Pixels per chip / ADCs per chip x Storage cells x bits per sample x (2 = analog and gain information) x trains per second.} $\approx$~700~Mbit/s. 


The digital PCBs handle the datastream of all 64 ADCs of the corresponding module, 45~Gbit/s in total. The core of the digital design is a VIRTEX-5 FPGA, which allows limited data sorting before transmitting it to the data acquisition (DAQ) system. The digital electronic PCB assembly is shown in figure \ref{digital_pcb}.

In addition to the pixel data, the bunch number associated with each storage cell and debug information for service purposes is transferred to the DAQ system.

One of the digital boards is designed as a functional mezzanine board, which is reused for other detector projects with high speed data readout such as LAMBDA \cite{lambda, lambda2} and PERCIVAL \cite{percival}. Therefore four 10~GbE outputs are integrated on the mezzanine, of which AGIPD uses only one. The 1 Megapixel AGIPD system will deliver 80~Gbit/s to the off-detector DAQ system on 16 links in total. Each link is a 10~GbE/UDP link and will be used with an average data rate of approximately 5~Gbit/s.

\subsection{Data Acquisition (DAQ) system}

The off-detector DAQ is developed as a common component for use with all detectors at European XFEL beamlines. For the 2D cameras and potentially other systems (digitizers, fast ADCs, etc.), an ATCA data acquisition train builder card \cite{coughlan} is being developed to receive, sort and reject data of cameras with 1~Megapixel or more. Data from two AGIPD modules is received by an input train builder FPGA which performs additional module specific data sorting before storing to DDR memory. Final sorting to full 1~Megapixel images and full trains is performed from the input memory to the DDR2 memory of an output FPGA. The latter then sends the bunch ordered images and additional data to a PC-farm for further processing and archiving.


\subsection{AGIPD mechanics}

\begin{figure}[tb!]
  \centering
 \includegraphics[width=0.45\textwidth]{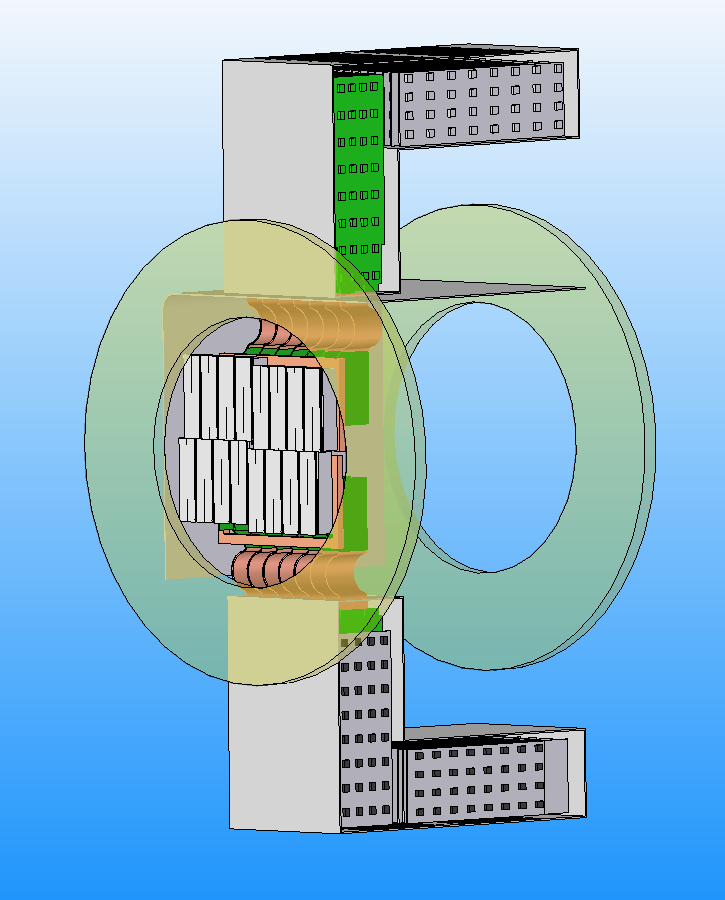} 
  \caption{Concept of the upstream detector of 1~Megapixel (16 modules).}
  \label{upstream}
\end{figure}

The basic idea of the mechanical concept is to separate the actual sensing elements (modules), which will be operated in the detector vacuum\footnote{From experience with experiments at the Linear Coherent Light Source (LCLS) it is recommended to separate the vacuum of the sample chamber from the detector vacuum in order to avoid the accumulation of sample residue on sensitive parts of the detector.}, from the interface electronics, which will be outside of the vacuum. 

In order to reduce the amount of space occupied along the beam axis, the electronics was designed in an angled shape, and form so called wings. Each wing will feature a closed loop air stream inside it to transport the heat dissipated in the interface electronics (about 500~W) to a special, water cooled heat exchanger in order to minimize the heat dissipated to the air of the experimental area.

\begin{figure}[tb!]
  \centering
 \includegraphics[width=0.45\textwidth]{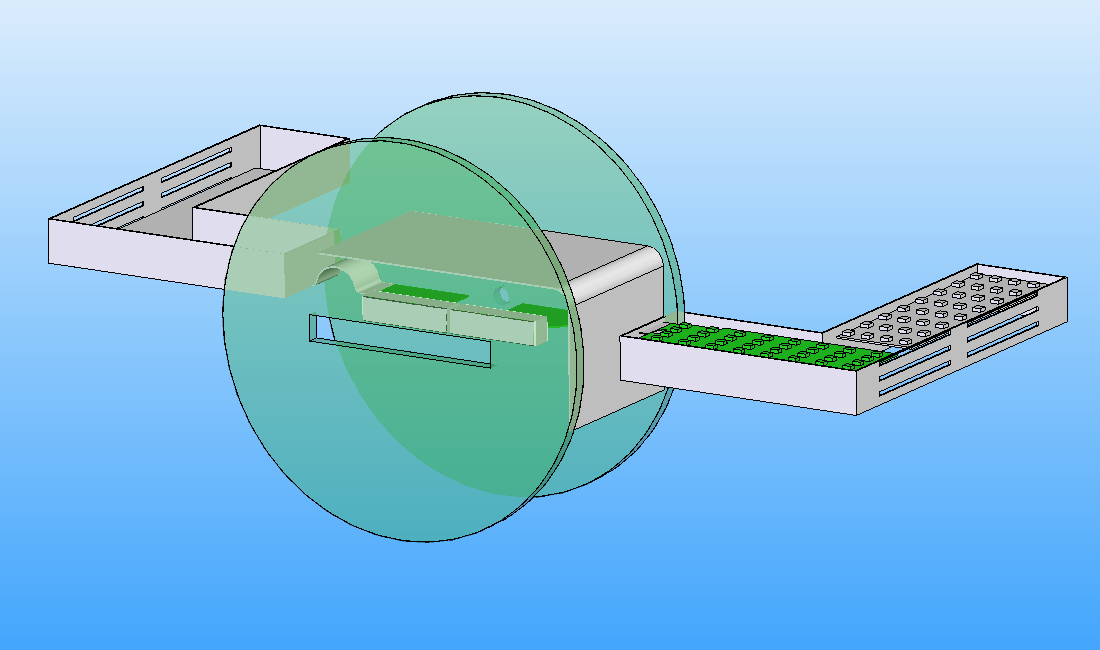}
  \caption{Concept of the downstream detector of 128k~pixels (2 modules).}
  \label{downstream}
\end{figure}

The detector will consist of one main (upstream) detector of 1~Megapixel, shown in Figure \ref{upstream}, which consists of 4 quadrants. In order to increase the angular coverage towards the central beam, the SPB beamline at the European XFEL will additionally employ a 128k downstream detector consisting of 2 modules, as shown in Figure \ref{downstream}.

To achieve the best possible scientific outcome the four quadrants of the main detector and the two modules of the downstream detector need to be movable, primarily to adjust the size of the central hole.

\subsubsection{Modules}

The basic building block of the detector system is the so called module, where each module is, in principle, an independent detector unit.

A so called hybrid is formed by an array of 2 x 8 ASICs which are bump bonded to the monolithic pixelated silicon sensor described above. The bump bonding will be performed at PSI using their well developed in-house technology \cite{psi_bonding}.

The hybrid is then glued onto a so-called sensor board made from low temperature co-fired ceramics (LTCC) material. The hybrid is then electrically connected to this LTCC board by wirebonding to form the bare module. Optionally a heatspreader of up to 500~$\upmu$m thickness can be mounted in between the hybrid and the sensor board to improve temperature homogeneity and to take up stresses from the mismatching thermal expansion coefficients of silicon and the LTCC material.

To complete the full module the bare module needs to be mounted onto the quadrant cooling block and the vacuum board needs to be connected.  

\subsubsection{Cooling system}
As the total power dissipation inside the vacuum will be around 1~kW, a cooling concept using a liquid coolant was adopted. The main coolant will be provided by a cooling plant outside the experimental hutch and has to be brought to, and extracted from, the experimental area via a dedicated pipe system. 

\begin{figure*}[tb!]
  \centering
 \includegraphics[width=0.95\textwidth]{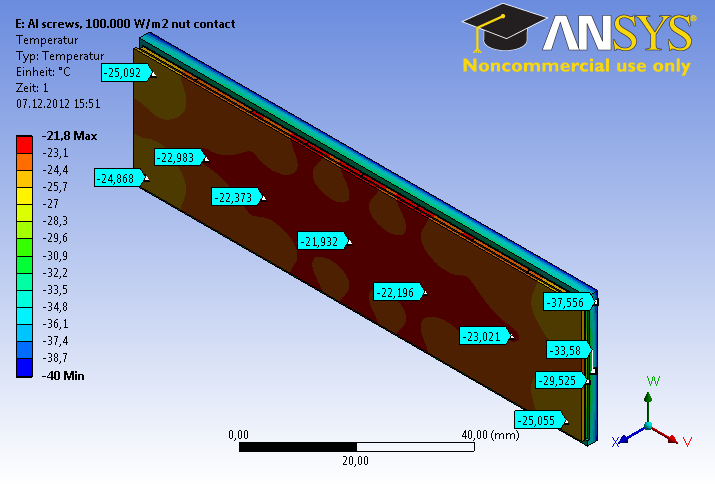}
  \caption{Thermal simulation of a bare module connected to the cooling block. The cooling block is at -40$^\circ$C, the warmest point at -21.9$^\circ$C. The temperature spread within the sensor is 3.5$^\circ$C.}
  \label{thermal_sim}
\end{figure*}

First the main coolant will flow into the four quadrant cooling blocks in parallel. The cooling blocks have internal cooling channels that meander through the available volume before leaving the quadrant cooling blocks. The outflowing coolant will then be fed into a secondary cooling block that, in turn, removes the heat from the voltage regulators on the vacuum board.

This concept is beneficial for the temperature uniformity of the ASICs, as it isolates the ASICs from the heat load of the vacuum boards. Additionally the voltage regulators only need to be stabilized in temperature; the temperature uniformity is of secondary importance.

As stated above, it is intended to operate irradiated ASICs at a temperature of -20$^\circ$C or below to reduce the droop on the analog storage cells. Simulations of the sensor stack (Figure \ref{thermal_sim}) show that the temperature difference between the hottest and the coldest point on the sensor is less than 4$^\circ$C, and that all ASICs operate at a temperature of -22$^\circ$C or lower when the main coolant is at a temperature of -40$^\circ$C. Slight variations of these values are expected, as the coolant temperature will increase while passing through the quadrant. The largest temperature drop, $> 10^\circ$C, happens in the 2.4 mm thick LTCC material of the sensor board.

In order to maximize the temperature stability of the cooling system, a chiller system using a bath design with a volume of more than 20~l was chosen. A polydimethyl siloxane (PDMS\footnote{PDMS is commonly known as silicone oil}) based coolant, which fulfills the necessary requirements of low viscosity and chemical inertness in the anticipated temperature range, will be used.

\section{Summary and outlook}

The AGIPD project, a joint detector development program by DESY, PSI and the universities of Hamburg and Bonn, is progressing and the first module is expected to be operational before the end of 2013. 

The test chips allowed the investigation of the performance of individual ASIC components, including their radiation hardness, and the results have proven invaluable for the design of the full scale chip AGIPD 1.0, the submission of which is immanent.

The full 1 Megapixel system will be deployed at the European XFEL in 2015 for commissioning and calibration and will be available for user operation at "Day 1" in 2016.

\end{document}